\documentclass[
  reprint,
  prl,
  superscriptaddress,
  amsmath,
  amssymb,
  floatfix
]{revtex4-2}
\usepackage{graphicx}% Include figure files
\usepackage{dcolumn}% Align table columns on decimal point
\usepackage{bm}% bold math
\usepackage{amsmath}
\usepackage{empheq}
\usepackage[nopatch]{microtype}
\usepackage{booktabs}
\usepackage{float}
\usepackage{graphicx}
\usepackage[utf8]{inputenc}
\usepackage{amsmath}
\usepackage{xcolor}
\usepackage{soul}
\usepackage{siunitx}
\usepackage{braket}
\usepackage{url}

\newcommand{\affUPB}{Department of Physics and Center for Optoelectronics and Photonics Paderborn (CeOPP), Paderborn University, Warburger Straße 100, 33098 Paderborn, Germany}
\newcommand{\affPhoQSonly}{Institute for Photonic Quantum Systems (PhoQS), Paderborn University, Warburger Str. 100, Paderborn, 33098, Germany}
\newcommand{\affKIT}{Institute of Photonics and Quantum Electronics (IPQ) and Center for Integrated Quantum Science and Technology (IQST), Karlsruhe Institute of Technology, Engesserstr. 5, 76131 Karlsruhe, Germany}
\newcommand{\affJKU}{Institute of Semiconductor and Solid State Physics, Johannes Kepler University Linz, Linz, Austria}
\newcommand{\affSapienza}{Department of Physics, Sapienza University of Rome, 00185 Rome, Italy}
\newcommand{\affJMU}{Technische Physik, Julius-Maximilians-University of W\"urzburg, Am Hubland, 97074 W\"urzburg, Germany}
\newcommand{\affFHVB}{Research Center for Microtechnology, Vorarlberg University of Applied Sciences, Campus V, Hochschulstrasse 1, 6850 Dornbirn, Austria}

\begin{document}

\title{Beyond Antibunching: Photon Correlation Analysis Reveals Blinking Origin}

\author{Patricia Kallert}
\thanks{These authors contributed equally to this work.}
\affiliation{\affUPB}
\affiliation{\affPhoQSonly}

\author{Santiago Bermúdez-Feijóo}
\thanks{These authors contributed equally to this work.}
\affiliation{\affUPB}
\affiliation{\affPhoQSonly}

\author{Giorgio De Pascalis}
\affiliation{\affUPB}
\affiliation{\affPhoQSonly}

\author{Nicolas Claro-Rodriguez}
\affiliation{\affUPB}
\affiliation{\affPhoQSonly}

\author{Ioannis Caltzidis}
\affiliation{\affUPB}
\affiliation{\affPhoQSonly}

\author{Normen Auler}
\affiliation{\affUPB}
\affiliation{\affPhoQSonly}

\author{Eva Berger}
\affiliation{\affUPB}
\affiliation{\affPhoQSonly}

\author{Rebecca Aschwanden}
\affiliation{\affUPB}
\affiliation{\affPhoQSonly}

\author{Tobias Krieger}
\affiliation{\affJKU}

\author{Saimon F. Covre da Silva}
\altaffiliation[Current address: ]{Instituto de Física Gleb Wataghin, Universidade Estadual de Campinas (UNICAMP), Campinas, Brazil}
\affiliation{\affJKU}

\author{Sandra Stroj}
\affiliation{\affFHVB}

\author{Quirin Buchinger}
\affiliation{\affJMU}

\author{Michele B. Rota}
\affiliation{\affSapienza}

\author{Thomas Hummel}
\affiliation{\affPhoQSonly}

\author{Sven Höfling}
\affiliation{\affJMU}

\author{Armando Rastelli}
\affiliation{\affJKU}

\author{Rinaldo Trotta}
\affiliation{\affSapienza}

\author{Dirk Reuter}
\affiliation{\affUPB}
\affiliation{\affPhoQSonly}

\author{Sonja Barkhofen}
\email{Corresponding author: sonja.barkhofen@upb.de}
\affiliation{\affUPB}
\affiliation{\affPhoQSonly}

\author{Klaus D. Jöns}
\affiliation{\affUPB}
\affiliation{\affPhoQSonly}

\author{Tobias Huber-Loyola}
\affiliation{\affKIT}

\date{\today}

\begin{abstract}

A variety of quantum emitters compete in the quest for the best single-photon source for photonic quantum technologies. Consequently, only a consistent approach to analyze the second-order auto-correlation function allows comparison of the multi-photon contribution ($g^{(2)}(0)$) of different sources. However, the community employs different and inconsistent methods for blinking sources, leading to incomparable benchmarks of source quality. Here, we use the emission of an inherently non-blinking quantum dot (QD) and apply artificial blinking through two different mechanisms: masking the recorded raw data in post-processing and emulation of a blinking system by gating the laser excitation pulses. We then compare the $g^{(2)}(0)$ values with the non-blinking result. For the analysis, we investigate five estimators of $g^{(2)}(0)$ actively used in the literature. While fitting the envelope of the correlations on long-time scales with the correct blinking model is the best choice, normalizing to the Poisson level gives by far the worst proximity. Furthermore, we test our predictive model to identify the underlying blinking mechanism of a QD in a circular Bragg grating cavity.
\end{abstract}


\maketitle

The second-order auto-correlation function, $g^{(2)}(\tau)$, is one of the most widely used observables in quantum optics. Since the seminal work of Glauber, photon correlations have provided a direct experimental route to distinguish coherent, thermal, and nonclassical light fields \cite{Glauber.1963}. In particular, the zero-delay value $g^{(2)}(0)$ has become the standard figure of merit for single-photon emission: values below unity indicate photon antibunching, while $g^{(2)}(0)<0.5$ is commonly used as the operational criterion for emission dominated by single photons. This simple number is therefore routinely used to benchmark quantum emitters, including atoms, molecules, color centers, and semiconductor quantum dots (QDs), which we will focus on in the following \cite{Esmann.2024}.

The normalized Glauber correlation function is defined as
\begin{equation}\label{eq:g2tau_glauber}
    g^{(2)}(\tau)=\frac{\langle \hat a^\dag (t) \hat a^\dag (t+\tau) \hat a(t+\tau) \hat a(t)\rangle}{\langle \hat a^\dag (t) \hat a(t) \rangle \langle \hat a^\dag (t+\tau) \hat a(t+\tau)\rangle},
\end{equation}
where $\hat a(t)(\hat a^\dag (t))$ is the photon annihilation (creation) operator for a single photon in a single optical mode at time $t$. For a stationary source, the average intensity is constant over an observation window much larger than the characteristic time scales of the sources' fluctuations. Thus, the denominator of Eq. (\ref{eq:g2tau_glauber}) reduces to the constant $ \langle \hat n \rangle ^2 = \langle \hat a^\dag \hat a\rangle^2$, which is the square of the expectation value of the photon-number operator $\hat n$. For pulsed excitation measurements, the autocorrelation data consists of a comb of peaks spaced by the laser repetition rate, with a distinctive dip at zero delay for a perfect single-photon source. The integrated area of each side peak, $S_k$, remains constant in the absence of long-lived correlations, providing a consistent level that can be used for the normalization of the $g^{(2)}(\tau)$, usually referred as the Poisson level \cite{Michler.2024}. The zero-delay value is then defined as the ratio of the integrated zero-delay peak, $A_0$, to the normalization with the integrated average of the side-peaks, $\langle S_k\rangle$, thus giving  $g^{(2)}(0)={A_0}/{\langle S_k\rangle}$. However, this simplification can fail when the mean photon flux varies during the measurement, leading to apparent changes in the value of $g^{(2)}(0)$, even when the underlying photon statistics during the bright events remain the same.

\begin{figure}[t]
    \centering
    \includegraphics[width=1\linewidth]{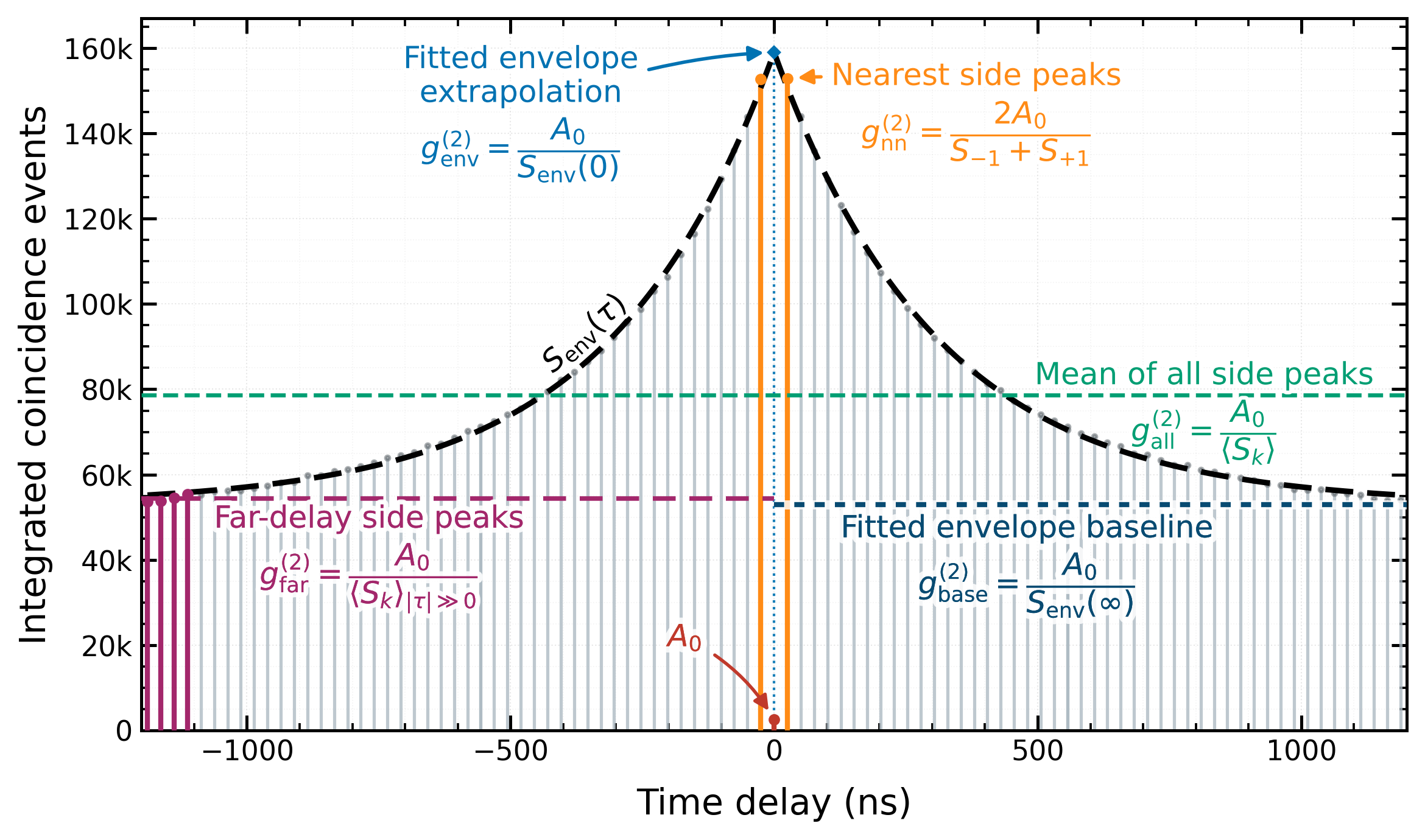}
    \caption{Normalization methods for estimating $g^{(2)}(0)$ from a pulsed correlation histogram. The central peak area $A_0$ is compared with different reference levels extracted from the side peaks according to Eqs. (\ref{eq:g2estimators}a-\ref{eq:g2estimators}e).}
    \label{figEstimators}
\end{figure}

A particularly relevant form of non-constant flux is blinking, i.e., stochastic switching between periods of emitting and dark states of the source \cite{Mason.1998}.
Blinking is ubiquitous in solid-state quantum emitters and may arise from long-lived charged \cite{Santori.2004} or metastable states \cite{Zhao.2020},  carrier trapping \cite{Davanco.2014}, local photo-ionization \cite{Efros.2016}, spectral diffusion \cite{Robinson.2000,Frantsuzov.2008}, surface charges \cite{Quinn.2016}, or changes in the excitation and collection conditions.  In a correlation measurement, blinking produces a modulation of the intensity correlations on timescales that can be orders of magnitude longer than the radiative lifetime, showing the characteristic decrease of the side peak counts. These slowly-varying correlations are physically real, but they leave room to misinterpret a different quantity that is often desired in single-photon-source characterization and that should be independent of these correlations: the intrinsic multi-photon emission probability of the emitter during its bright state. As a result, the extracted $g^{(2)}(0)$ can depend on the chosen normalization window, the acquisition time, the blinking statistics, and the treatment of dark periods. This is especially problematic when $g^{(2)}(0)$ values derived from different parameter choices are used as a benchmark number for comparing different emitters or devices' quality. Additionally, it is especially relevant in experiments involving remote, blinking QDs and other single-photon sources, since multi-photon contribution and the blinking statistics are used to correct, for instance, their interference visibility  \cite{Gold.2014,Somaschi.2015,Portalupi.2016,Thoma.2017,Reindl.2017,Vural.2018,Weber.2018}. Under blinking conditions, the side peak area is no longer independent of delay, but follows the slowly varying envelope $S_{\mathrm{env}}(\tau)$. As a result, different choices of the normalization level select specific features of the blinking envelope, and therefore, lead to different operational values of $g^{(2)}(0)$. Motivated by this, we compare five estimators that are visualized in Fig. \ref{figEstimators},
\begin{subequations}\label{eq:g2estimators}
\begin{align}
g^{(2)}_\mathrm{all}(0) &= \frac{A_0}{\left\langle S_k\right\rangle},\\
g^{(2)}_\mathrm{nn}(0) &= \frac{2 A_0}{S_{-1}+S_{+1}},\\
g^{(2)}_\mathrm{far}(0) &= \frac{A_0}{\left\langle S_k\right\rangle_{|\tau|\gg 0}},\\
g^{(2)}_\mathrm{env}(0) &= \frac{A_0}{S_\mathrm{env}(0)},\\
g^{(2)}_\mathrm{base}(0) &= \frac{A_0}{S_\mathrm{env}(\infty)}.
\end{align}
\end{subequations}
Here, $g^{(2)}_\mathrm{all}$ corresponds to the common procedure of averaging all selected side peaks in the analyzed correlation window. This estimator does not assume a blinking model, and uses side peaks taken at different delays, being therefore sensitive to the chosen time window when the side peak envelope is not flat \cite{Gazzano.2013,Fischer.2017,Schweickert.2018,Scholl.2019,Hauser.2026}. The estimator $g^{(2)}_\mathrm{nn}$ uses as a normalization only the integrated mean of the two nearest neighboring side peaks ($S_{\pm 1}$) to the central delay $A_0$ peak \cite{Santori.2002, Neuwirth.2022,Kaupp.2023,Ma.2026b}. This estimator is also obtained directly from the measured peak areas.
The estimator $g^{(2)}_\mathrm{far}(0)$ normalizes with respect to the mean value of the long time-delayed side peaks. Thus, it tries to evaluate the Poisson level of the auto-correlation function by using the raw data, without involving fitting models. Usually, the here considered peaks are the furthest delayed in the displayed histogram. The $g^{(2)}_\mathrm{env}$ estimator uses the fitted side peak envelope to extrapolate the appropriate reference level to zero delay, $S_{\mathrm{env}}(0)$. This is the model-based estimator that corrects for the slow-blinking envelope. If blinking is the only source for the vacuum contributions, this estimator corresponds to the $g^{(2)}_\mathrm{G}(0)$ definition introduced by Grünwald in 2019 \cite{Grunwald.2019}, which we explain in detail in the appendix. Finally, $g^{(2)}_\mathrm{base}$ normalizes the central peak to the far-delay level $S_{\mathrm{env}}(\infty)=C$, where the blinking-induced bunching has decayed and the envelope reaches a baseline \cite{Santori.2001,Santori.2004,Davanco.2014,Miyazawa.2016,Karli.2024,Pennacchietti.2024,Thomas.2024,Yang.2024,Joos.2024,Wijitpatima.2024,Mudi.2026,Aschwanden.2026}.\\\\

To understand how different parameter choices, like blinking time scales or correlation window width, impact the extracted $g^{(2)}(0)$ estimators and how to identify the correct normalization procedure, we apply a systematic approach.
As a reference, we use an epitaxial semiconductor quantum dot transition as an inherently non-blinking two-level system. We perform a Hanbury Brown and Twiss experiment \cite{Brown.1956,Kimble.1977} that is the established method for measuring the second-order correlations to characterize QDs as single-photon sources \cite{Michler.2000}. Based on this, we generate a well-defined and known blinking pattern by applying two complementing modifications. First, we temporally mask the raw time-tagged photon streams and autocorrelate the resulting new data. Second, the excitation pulse is gated by a pulse picker, controlled by a signal from an arbitrary waveform generator (AWG). Based on these methods, we provide a practical framework for consistently comparing single-photon sources in the presence of blinking.

\begin{figure}[t]
    \centering
    \includegraphics[width=1\linewidth]{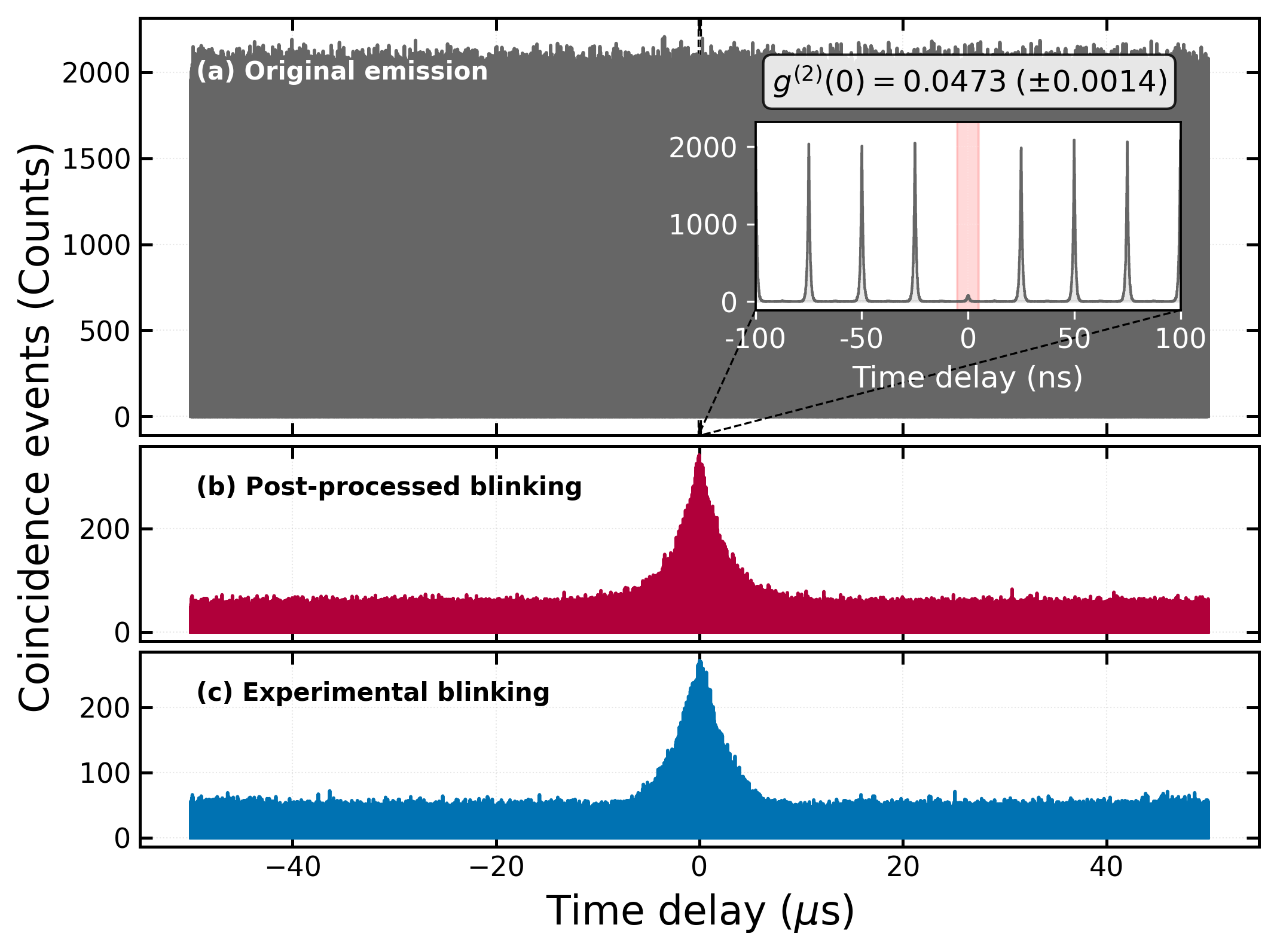}
    \caption{Correlation histograms.
    (a) Reference measurement of the initially non-blinking quantum dot emission, where a flat side peak envelope is obtained for long time delays. The inset displays the short-time pulsed correlation. A value of $g^{(2)}(0)=\num{0.0473\pm0.0014}$ is extracted using Eq. (\ref{eq:g2estimators}a).
    (b) Artificial blinking generated by timestamp masking, applying an RTN waveform to the raw time-tagged photon streams. The imposed intermittency produces a pronounced rapidly decaying envelope in the side peak amplitudes with increasing delay, here shown for $r=P_{\mathrm{off}}/P_{\mathrm{on}}=\num{5.3}$ and $\tau_b=\qty{2.3}{\micro\second}$.
    (c) Experimentally induced blinking obtained by gating the excitation pulse train with an RTN waveform using a pulse picker. Blinking ratio $r=5.287\pm0.015$ and timescale $\tau_b = \qty{2.309\pm0.08}{\micro\second}$ are extracted from fitting the integrated peaks with the RTN envelope given by Eq. (\ref{eq:S_env_RTN}) in the appendix.}
\label{figCorrelationPlots}
\end{figure}
The recorded coincidence histogram of the non-blinking emission from our QD, shown in Fig. \ref{figCorrelationPlots}(a), exhibits a flat side peak envelope and yields a reference antibunching value of $g^{(2)}_\mathrm{all}(0)$ = 0.0473 ± 0.0014, providing the baseline photon statistics against which all estimators are compared. In this work, we use Gaussian error propagation with Poissonian statistics for the uncertainty evaluations. Furthermore, we send pulses from an optical parametric oscillator (OPO) through a pulse picker and slice them with a 4-f pulse slicer to control the excitation bandwidth. We use $\pi$-pulse resonant excitation of the trion transition in an InAs QD with near Fourier-transform limited emission, integrated in a p-i-n diode for charge control \cite{Kuhlmann.2015} and further characterized in Fig. \ref{fig:QDCharacterization} in the appendix.

For the first approach, blinking is imposed in post-processing by applying a temporal mask to both raw time-tagged data streams recorded by the two detectors of the autocorrelation setup of the unblinking source \cite{Brown.1956}. The mask defines bright (ON) and dark (OFF) periods that follow a stochastic binary waveform from one of three blinking mechanisms: Random Telegraph Noise (RTN) \cite{Efros.1997,Pistol.1999}, Dark States (DS) \cite{Davanco.2014} and Powerlaw (PL) processes, which affect, for instance, colloidal quantum dots \cite{Lounis.2000,Shimizu.2001,Kuno.2001,Frantsuzov.2008,Nguyen.2013,Rabouw.2019}. The blinking strength is controlled through the intermittency ratio $r$ = $P_{\mathrm{off}}/P_{\mathrm{on}}$ for each generated waveform. $P_{\mathrm{off}}$ and $P_{\mathrm{on}}$ denote the probabilities of the QD being emitting or not emitting, respectively. The two time-filtered data streams from the detectors are used to build the blinking modified coincidence post-processed histogram. An example of a  blinking histogram from this method for an RTN mechanism is shown in Fig. \ref{figCorrelationPlots}(b), for a waveform with a ratio of $r$ = 5.3 and a timescale of $\tau_b = \qty{2.3}{\micro\second}$, which represents the mean time of the QD emitting photons.

In the second approach, blinking is experimentally imposed on the excitation. An AWG produces a waveform that follows an RTN signal, which is employed as an external TTL trigger for an acousto-optic pulse picker. Depending on the instantaneous ON/OFF state of the waveform, the pulse picker either transmits or suppresses pulses from the excitation train and, therefore, creates physical dark periods. We record the coincidence histograms between the two outputs of the autocorrelation setup. The blinking parameters can then be extracted from fitting the integrated peaks with the RTN envelope given by Eq.~(\ref{eq:S_env_RTN}) in the appendix. An example from this approach is shown in Fig.  \ref{figCorrelationPlots}(c) with an intermittency ratio of $r=5.287\pm0.015$ and a blinking timescale $\tau_b = \qty{2.309\pm0.08}{\micro\second}$.

\begin{figure}[t]
    \centering
    \includegraphics[width=1\linewidth]{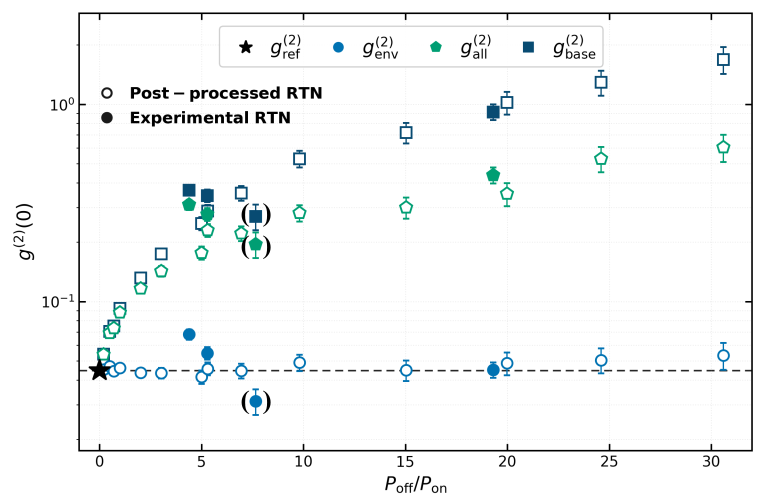}
\caption{$g^{(2)}(0)$ estimators for random telegraph noise blinking as a function of the blinking strength $P_{\mathrm{off}}/P_{\mathrm{on}}$. The non-blinking reference value is shown by the black star and dashed horizontal line. Open markers correspond to post-processed RTN blinking, while filled markers show experimentally induced RTN blinking. The data points within the bracket deviate from the expected trend due to a different pulsed suppression. The envelope and baseline estimators, $g^{(2)}_{\mathrm{env}}$ and $g^{(2)}_{\mathrm{base}}$, are obtained from the fit to the side peak envelope $S_\mathrm{env}(\tau)$ given by Eq. (\ref{eq:S_env_RTN}) in the appendix.}
\label{fig:g2RTNblinking}
\end{figure}

Based on the two aforementioned controllable blinking methods, we compare the estimators defined in Eq. (\ref{eq:g2estimators}) for an RTN process, as a function of the intermittency ratio $r$. The results are shown in Fig.\ref{fig:g2RTNblinking}, where 
we focus on the model based estimators $g^{(2)}_\mathrm{env}(0)$, $g^{(2)}_\mathrm{base}(0)$ and $g^{(2)}_\mathrm{all}(0)$, which is the most common approach for non-blinking QDs. The envelope estimator $g^{(2)}_\mathrm{env}(0)$ remains close to the reference value in the explored range of $r$ for both artificial blinking methods, while $g^{(2)}_\mathrm{all}$ and $g^{(2)}_\mathrm{base}$ lead to a significant overestimation of $g^{(2)}(0)$ with increasing blinking strength. $g^{(2)}_\mathrm{base}$, which normalizes to the Poisson level, even creates a $g^{(2)}_\mathrm{base}(0)>1$ that would imply bunched light emission for $P_{\mathrm{off}}/P_{\mathrm{on}}\geq 20$. For further insight into the shape of such correlation measurements, consult the appendix. The remaining deviations between the experimentally gated and timestamp-masked datasets are attributed to practical limitations of the pulse-gating implementation, including imperfect laser suppression (specified for data points within brackets in Fig. \ref{fig:g2RTNblinking}), residual leakage of the original $80$ MHz laser repetition rate, and uncertainty in the excitation power due to the non-homogeneous signal.

\begin{figure}[b]
    \centering
    \includegraphics[width=1\linewidth]{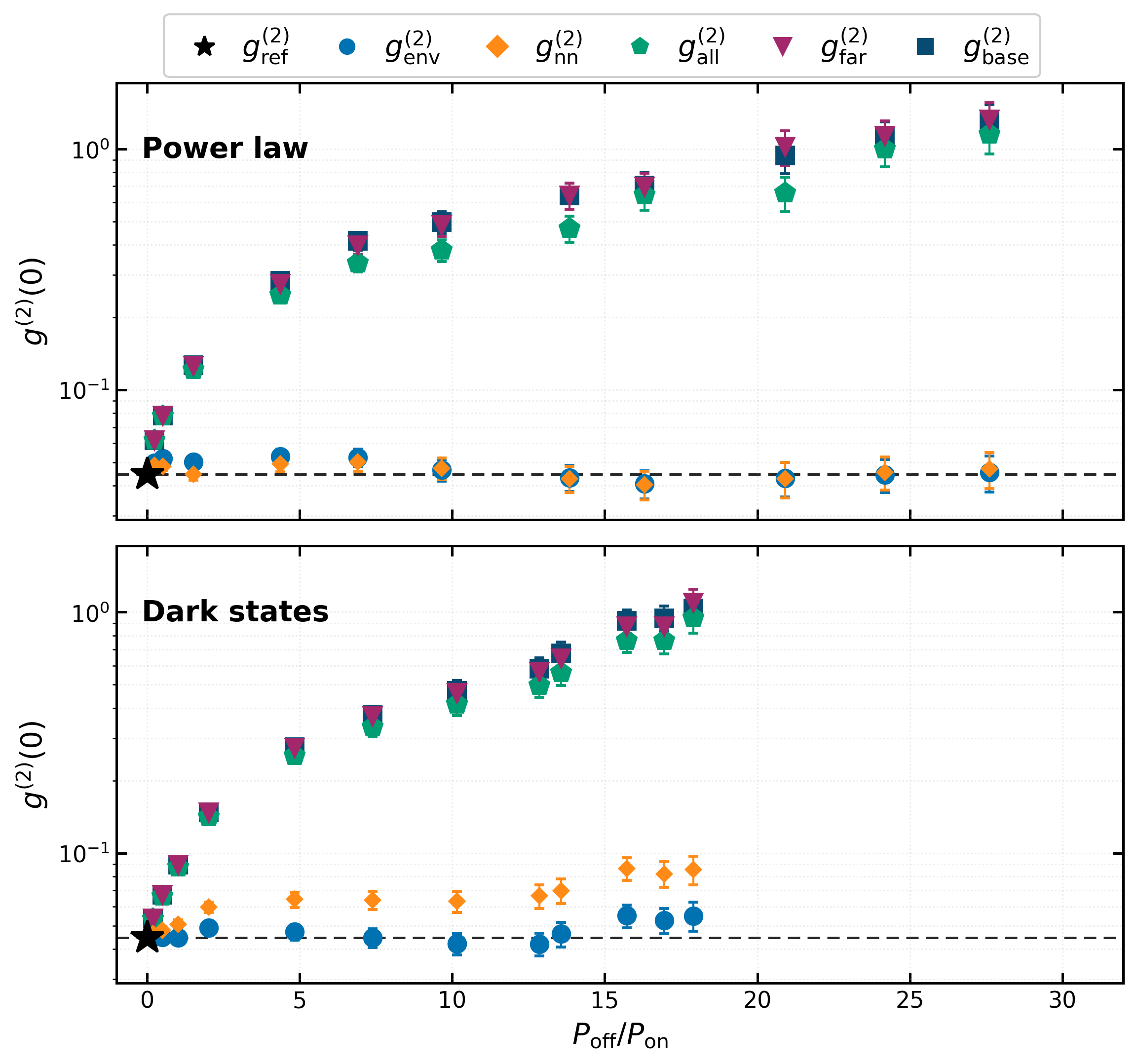}
    \caption{$g^{(2)}(0)$ estimators for power law and dark state blinking generated by post-processed blinking, shown as a function of the blinking strength $P_{\mathrm{off}}/P_{\mathrm{on}}$. These models are defined in the appendix. The non-blinking reference value is indicated by the black star and dashed horizontal line.}
\label{fig:g2blinkingDS_PL}
\end{figure}

Fig. \ref{fig:g2blinkingDS_PL}, on the other hand, shows the comparison solely based on the post-processing masking for the Powerlaw and Dark States blinking mechanisms. For the underlying physical processes, refer to the appendix. The non-blinking reference value of $g^{(2)}(0)$ is shown by a black star. The envelope and baseline estimators, $g^{(2)}_{\mathrm{env}}$ and $g^{(2)}_{\mathrm{base}}$, are obtained from the fit to the side peak envelope $S_\mathrm{env}(\tau)$ given by Eq. (\ref{eq:S_env_DS}, \ref{eq:S_env_PL}) in the appendix, while $g^{(2)}_{\mathrm{nn}}$, $g^{(2)}_{\mathrm{far}}$ and $g^{(2)}_{\mathrm{all}}$ are obtained directly from the measured side peak areas. Estimators that use delayed or averaged reference levels, particularly $g^{(2)}_\mathrm{base}$ and $g^{(2)}_\mathrm{all}$,  increase strongly with larger $r$. This drift does not reflect a change in the underlying single-photon emission, which remains fixed in the timestamp-masked data, but rather arises from the chosen normalization level. This dependency is particularly intuitive since blinking affects single- and multi-photon contributions simultaneously, which is more visible over long delays ($\tau\gg\tau_b$) due to the more prominent contribution of vacuum components. In contrast, $g^{(2)}_\mathrm{env}$ remains close to the non-blinking reference because it normalizes the central peak to the peak level expected at zero delay and removes the additional vacuum components \cite{Grunwald.2019}, described in detail in the appendix. Moreover, the nearest-side estimator $g^{(2)}_\mathrm{nn}$ follows $g^{(2)}_\mathrm{env}$ most closely, making it a useful model-independent approximation when the envelope fit cannot be reliably determined. The same qualitative hierarchy is observed for RTN, DS, and PL mechanisms, indicating that the estimator drift is not tied to a particular blinking model. Still, for DS, or in general very rapid blinking, the nearest neighbor estimator starts to differ with a systematic deviation from the original $g^{(2)}(0)$ and $g^{(2)}_\mathrm{env}$ for higher $P_{\mathrm{off}}/P_{\mathrm{on}}$ ratios and reaches its limits. The data from the experimentally induced blinking follows the same qualitative behavior as the timestamp-masked RTN data, supporting the observed estimator dependence, and showing that it is not an artifact of the post-processing procedure. We emphasize that for high $P_{\mathrm{off}}/P_{\mathrm{on}}$ ratios, the DS model reaches its limit in post-processing, since the coincidences then reduce significantly and are therefore not usable.\\\\

As a practical example, and to demonstrate how to apply the method to experimental data, we analyze the emission of an inherently blinking QD. In this case, we use a GaAs/AlGaAs quantum dot obtained by droplet etching epitaxy, integrated in a Circular Bragg Grating resonator, as used in \cite{Aschwanden.2026}. We use the neutral exciton transition under phonon-assisted, pulsed excitation, with an auxiliary weak green continuous-wave laser to provide a degree of charge stabilization \cite{Yang.2022}. All unwanted laser contribution is removed with notch filters, as well as a transmission monochromator, both in the wavelength range of 795 nm. Further details on the experimental setup can be found in the appendix. 

\begin{figure}[h]
    \centering
    \includegraphics[width=1\linewidth]{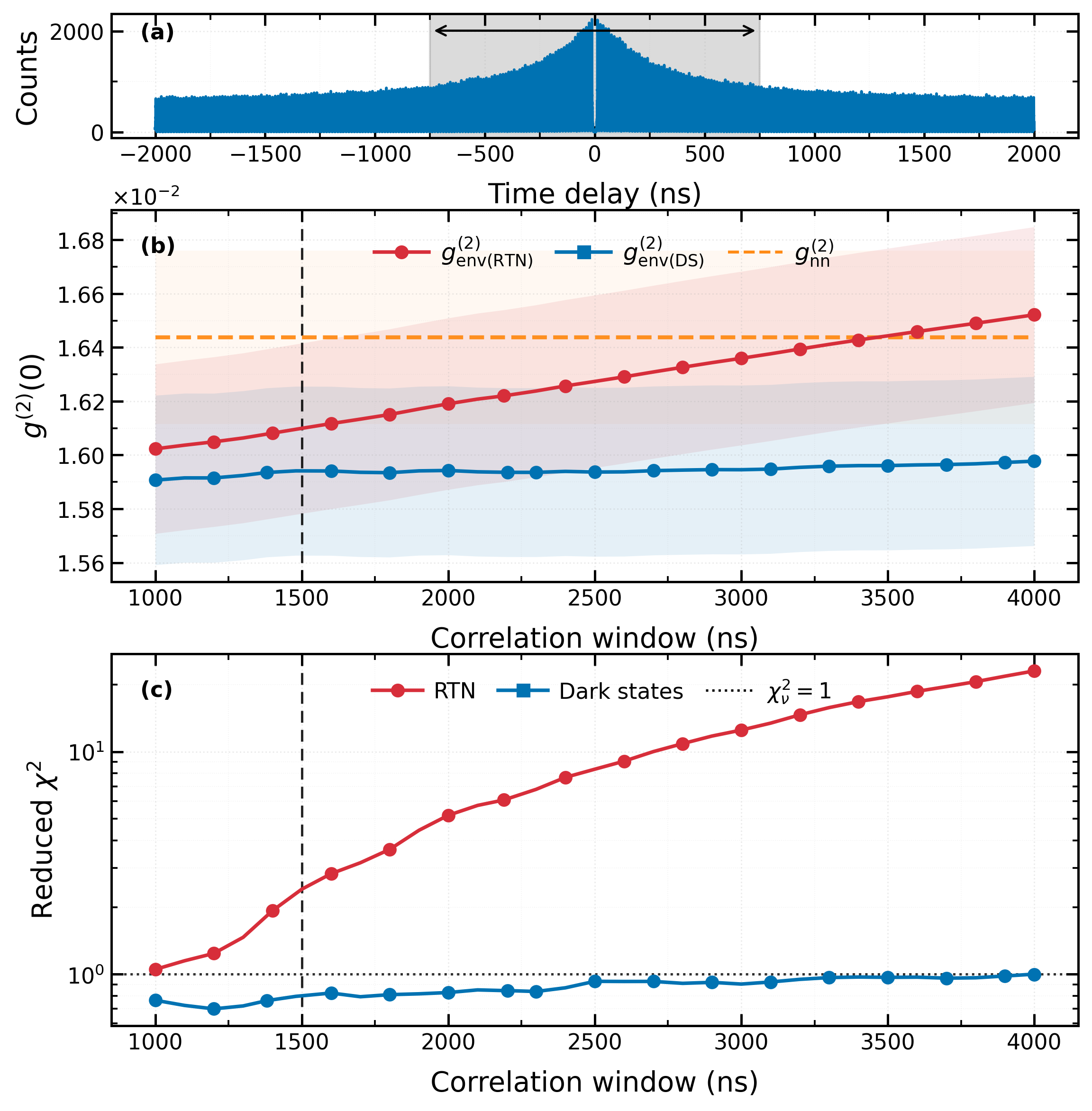}
\caption{$g^{(2)}(0)$ estimators for an inherently blinking quantum dot under phonon-assisted excitation. (a) Measured pulsed correlation histogram showing a pronounced bunching envelope due to blinking. The shaded region indicates the correlation window used for the estimator calculation, with the arrows marking the corresponding window width. (b) Extracted $g^{(2)}(0)$ estimator values as a function of the correlation window. The vertical dashed line corresponds to the window shown in panel (a). (c) Reduced $\chi ^2$ for increasing correlation window for both the fit with random telegraph noise blinking and the dark state model.}
\label{fig:Analysis_NaturallyBlinkingQD}
\end{figure}

The second-order autocorrelation function from the naturally blinking transition is shown in Fig. \ref{fig:Analysis_NaturallyBlinkingQD}(a) (The same as used in Ref. \cite{Aschwanden.2026}, Fig. 3(b)).
In Fig. \ref{fig:Analysis_NaturallyBlinkingQD}(b), we evaluate $g^{(2)}(0)$ 
for two possible dominating blinking effects: one case uses the envelope for an RTN process, $g^{(2)}_{\text{env}(\mathrm{RTN})}$, and the other for a DS process, $g^{(2)}_{\text{env}(\mathrm{DS})}$. We vary the time window in which the correlation is analyzed, and compare the estimators with the model-free $g^{(2)}_\mathrm{nn}(0)$. From the previous results, this estimator proved to be the best model-free choice to determine the $g^{(2)}(0)$ value of a blinking emission. Thus, this constitutes our reference for the rest of the discussion. From our results, all estimator measurements are contained within the uncertainty window of $\pm$ 0.0003. The nearest neighbor estimator behaves independently of the window size, displaying a value $g^{(2)}_\mathrm{nn}(0) = 0.0164$. For the RTN, at short time scales the value predicted is lower than the nearest neighbor. But for long time scales, it reaches our reference value within the experimental uncertainty. In contrast, a fit with the Dark State model envelope (see Eq. (\ref{eq:S_env_DS})) results in a constant $g^{(2)}_{\text{env}(\mathrm{DS})}(0)=0.0159$, which is consistently below the nearest neighbor approach with still overlapping uncertainty. This agrees with the findings given in Fig. \ref{fig:g2blinkingDS_PL}. However, this conclusion shows that RTN is not the dominant source of blinking, while DS proves to be a much more accurate description of the long time scale decay of correlations according to the reduced $\chi^2$ in Fig. \ref{fig:Analysis_NaturallyBlinkingQD}(c). This applies especially as the correlation window increases, approaching a reduced $\chi^2=1$, while for the RTN fit it instead grows monotonically. We can attribute this to a mismatch in the model prediction and not to an underestimation of the experimental uncertainties. The necessity to not only consider the correct blinking mechanism, but also the limitations of the nearest neighbor estimator are emphasized through these insights. Despite this, the latter ($g^{(2)}_\mathrm{nn}$) establishes itself to still be the second-best, most practical choice and even better than a random choice of fit, if the underlying blinking mechanism is not easy to determine. Additionally, multiple blinking mechanisms that would require more complex descriptions cannot be excluded, but the fit with the DS model already converges to a reduced $\chi^2=1$. Furthermore, we emphasize that the nearest neighbor estimator is only applicable to pulsed-excitation blinking. Therefore, for continuous-wave schemes that exhibit blinking photon sources, selecting the correct fitting model is essential for accurately estimating the source's single-photon purity.
\\

Our findings show that the evaluation of $g^{(2)}(0)$ of blinking sources is - though frequently used - much more delicate than assumed at first glance. We provide a detailed comparison of the established methods, discuss their limits and present a recipe how to determine the most realistic value. Our pulse-gating and post-process masking approaches allow for the direct comparison of a blinking emission from various models of choice with the intrinsic non-blinking $g^{(2)}(0)$. Introducing blinking to this kind of source in a controllable way permits one to systematically study the estimators' role in the evaluation of the single-photon purity. While a normalization to the Poisson level systematically alters and overestimates the result, the envelope fit with the correct model gives the best approximation to the actual, non-blinking $g^{(2)}(0)$. The direct analysis of an inherently blinking quantum dot emission further demonstrates the importance of the choice of model.

A slightly less rigorous, but highly practical approach, is the normalization of the central-to-nearest neighbor side peaks. However, given that this estimator reaches its limits for fast blinking processes, the correct normalization can ideally be found from the fitted envelope evaluated at zero time delay, allowing to compensate for any vacuum contribution from blinking.
\nocite{*}

\section{\label{sec:level4}Acknowledgements}
This work was supported by the European Research Council (ERC) under grant LiNQs (101042672), by the Deutsche Forschungsgemeinschaft (DFG) through the Transregional Collaborative Research Centre TRR 142/3-2022 (231447078), and by the German Federal Ministry of Research, Technology and Space (BMFTR) through Qecs (13N16272) and TUF-ToPiQC (13N17238). Additionally, this work received support from the Austrian Science Fund (FWF) via the Research Group FG5 (10.55776/FG5) and the Cluster of Excellence quantA [10.55776/COE1], as well as from the Linz Institute of Technology (LIT) via the project PEPSI (LIT-2025-14-YOU-121), supported by the State of Upper Austria and the Austrian Federal Ministry of Education, Science and Research. This project has further received funding from the European Union's EIC Pathfinder Challenges Action (grant agreement No. 101115575), from the QuantERA II programme (Grant Agreement No. 101017733) via the projects QD-E-QKD and MEEDGARD (FFG Grant Nos. 891366 and 906046), from the European Union's Horizon Europe research and innovation programme under the EPIQUE Project (GA No. 101135288), and from MUR (Ministero dell'Università e della Ricerca) through the PNRR MUR project PE0000023-NQSTI.

We thank Lukas Hanschke for the processing of the InAs/GaAs QD diode. K.D.J. thanks Peter Grünwald for insightful email exchanges in 2018 that planted an idea that, years later, grew into the analysis presented here.
\bibliographystyle{unsrt}
\bibliography{References}

\clearpage

\begin{widetext}

\appendix

\section{Experimental setup and quantum dot characterization}\label{SetupCharacterization}
\begin{figure}[h]
    \centering
    \includegraphics[width=0.6\linewidth]{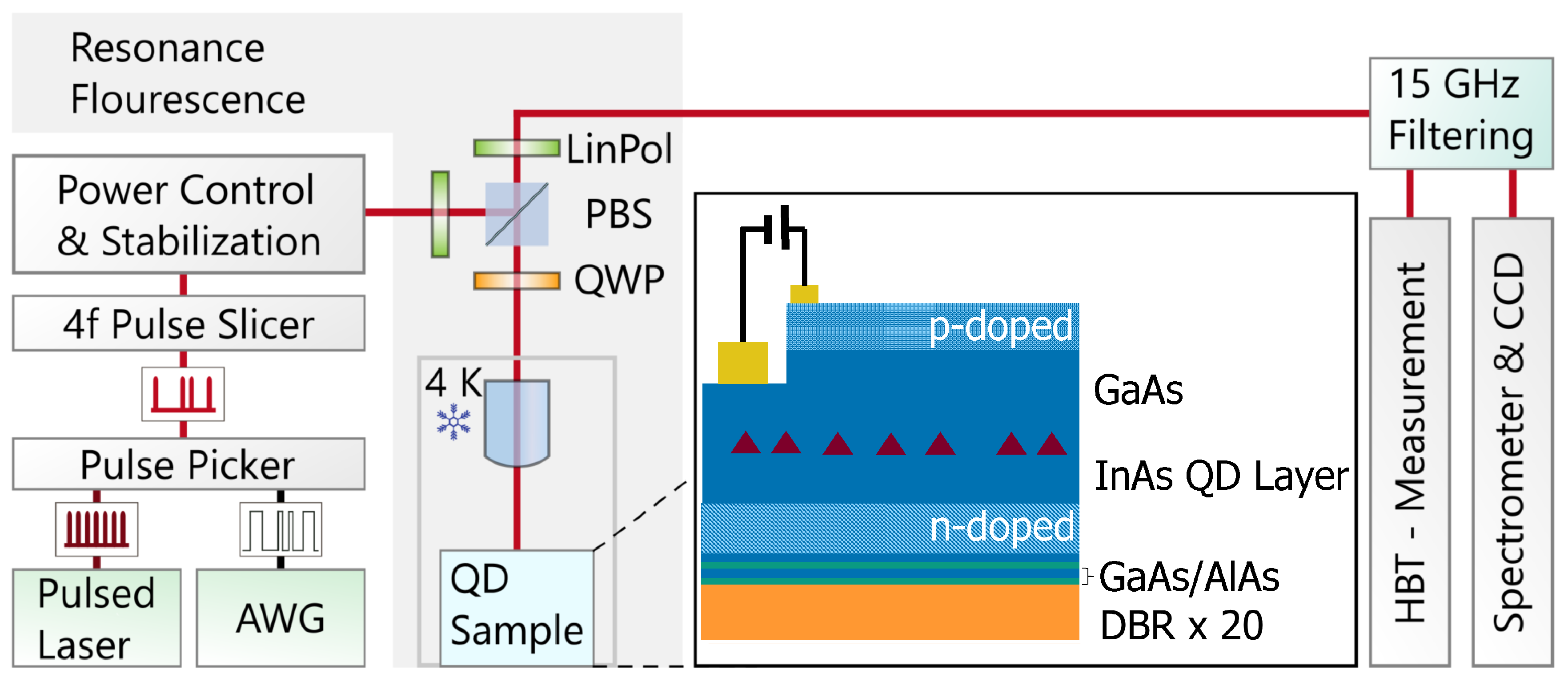}
\caption{Experimental Setup. A pulsed laser generates optical pulses with 80 MHz repetition and a pulse duration of 12 ps and is compatible with the AWG specifications. The AWG gates the pulses via a pulse picker, reducing the repetition to 40 MHz or applying the noise masking that can be programmed in the AWG. The pulses are sent then through a 80 m polarization-maintaining optical fiber from one lab to the other. The averaged power is stabilized by PID control and we excite the QD transition resonantly by filtering the exciting polarization of the laser with polarization suppression \cite{Kuhlmann.2013b}. The signal from the QD is filtered by a transmission monochromator of 15 GHz FWHM and then characterized by a spectrometer with CCD detector or the according Hanbury Brown \& Twiss experiment. The QD sample consists of a molecular beam epitaxy grown layer of Stranski-Krastanow InAs QDs in a GaAs matrix, while a p-doped and a n-doped layer create a diode structure with the QDs in the instrinsic layer. Distributed Bragg Reflectors (DBR) on the bottom enhance the extraction efficiency.}
\label{fig:ExperimentalSetup}
\end{figure}
We describe our setup in Fig. \ref{fig:ExperimentalSetup}, consisting of four major parts: the generation of the gated optical pulses from an Optical Parametric Oscillator (OPO) in one lab, the slicing and power-stabilization of the exciting pulses, the excitation and precharacterization of the QD transition and the filtering for the measurement of the second order correlation function as a Hanbury Brown and Twiss experiment.

As a source for our experiments we use InAs/GaAs quantum dots, grown by molecular beam epitaxy in Stranski-Krastanov mode. With a power-dependent photoluminescence measurement in Fig. \ref{fig:QDCharacterization}(a) we determine the power at which we receive the most effective population inversion from the ground to the excited state of the characterized two-level system. For the 40 MHz repetition, this is 91 nW with an additional neutral density filter of approximately 1 order of magnitude attenuation. This is needed to allow for decent stabilization even for high OFF/ON ratios. The diode integration allows us to excite the non-blinking negative trion transition resonantly in a dark field microscopy setup \cite{Kuhlmann.2013b}. To find the right voltage, we perform the voltage-dependent photoluminescence measurement in Fig. \ref{fig:QDCharacterization}(b) while exciting the QD with a 880 nm continuous-wave (CW) laser. At 730 mV, we operate the QD on the according Coulomb blockade plateau \cite{Faraon.2008}.

For a basic evaluation of the QD emission, we determine the lifetime $\tau_{E}$ of the transition from the convolution in Eq. (\ref{eq:RadiativeLifetime}a-c) as a correlation $c(t)$ between a trigger from the exciting pulse and the following detection of emitted photons. The according histogram is shown in Fig. \ref{fig:FTL}(a). Here, $y_0$ represents the background, $f_{populating}(t)$ ($f_{decay}(t)$) the function for the population (decay). $w$ is further the width of the Gaussian, $t_c$ its center and $a$ the area of the exponential function. Based on the measurements in Fig. \ref{fig:FTL}, we determine the QD emission to be close to the Fourier-transform limit, with a radiative lifetime $\tau_{E}$ = (649.94$\pm$0.57) ps and a linewidth $\nu$ (Eq. (\ref{eq:Lorentzian})) of 2.5 µeV that is determined by the emission response $I(\Delta E)$ to a weak, tuned and narrow-band (kHz regime) laser \cite{Gazzano.2018} and is displayed in Fig. \ref{fig:FTL}(b). Here, $\alpha$ is the area and $\Delta E$ the detuning from the center energy $E_{center}$.

For the inherently blinking QD we imply some modifications. Instead of the OPO we use a sliced fs MIRA with 80 MHz repetition and instead of the polarization suppression a 90 \% transmission and 10 \% reflection beam splitter as well as notches and another transmission monochromator for the according wavelength of the GaAs QDs in the circular bragg grating \cite{Rota.2024}.
\begin{subequations}\label{eq:RadiativeLifetime}
\begin{align}
    c(t) = y_0 + (f_{populating}*f_{decay})(t)\\
    f_{populating}(t)=\frac{1}{\sqrt{2\pi} w}e^{-\frac{(t-t_c)^2}{2w^2}}\\
    f_{decay}(t)=\frac{a}{\tau_{E}}e^{-\frac{t}{\tau_{E}}}
\end{align}
\end{subequations}
\begin{equation}\label{eq:Lorentzian}
 I(\Delta E)=I_0+\frac{2\alpha}{\pi}\Biggl(\frac{\nu}{4(\Delta E-E_{center})^2+\Delta E^2)}\Biggl)
\end{equation}
\begin{figure}[H]
    \centering
    \includegraphics[width=0.8\linewidth]{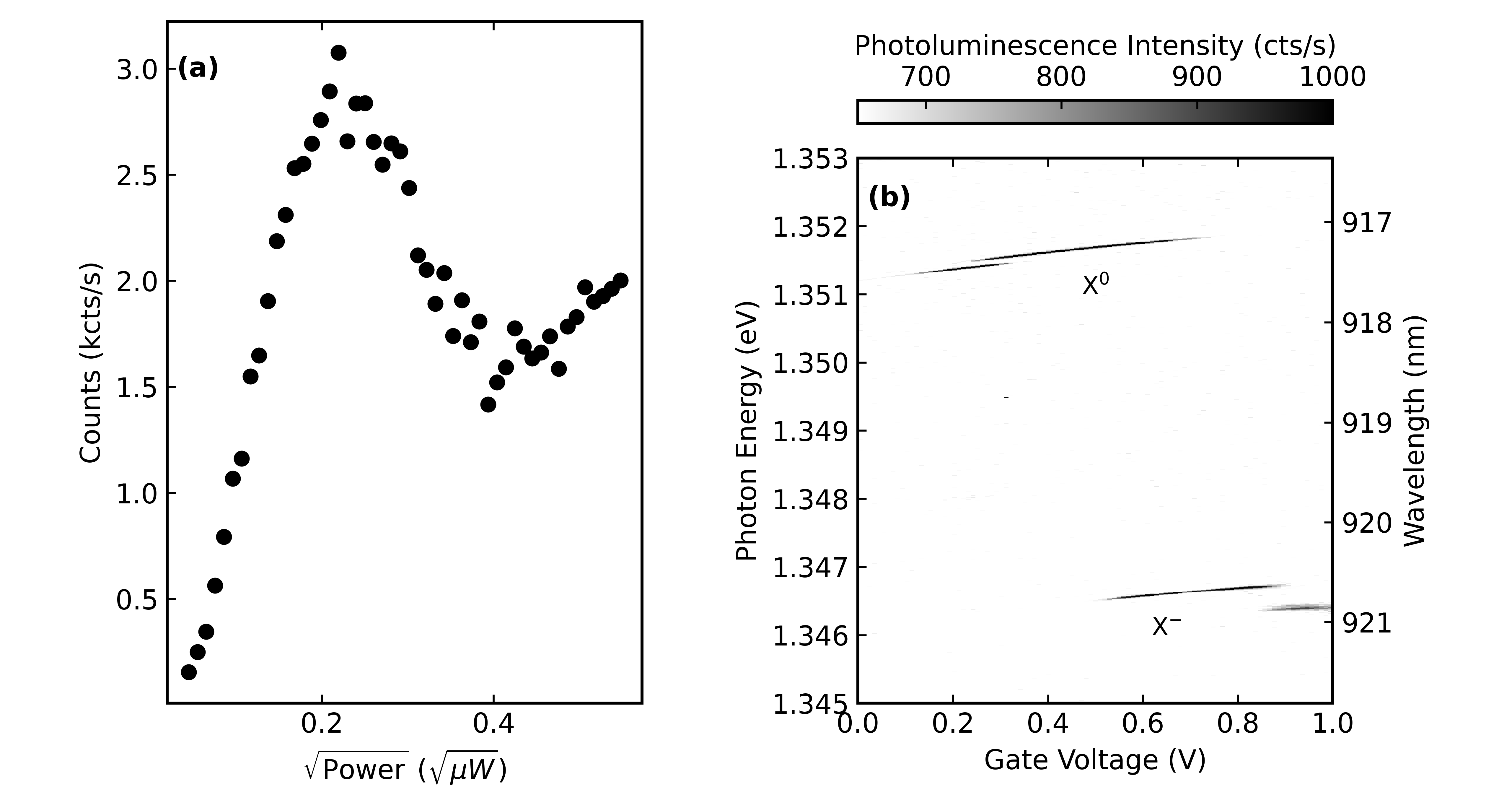}
    \hfill
\caption{(a) Used power-dependent photoluminescence (PL) signal in pulsed resonant excitation for 40 MHz repetition, i.e. for the non-masked excitation. (b) Voltage-dependent PL signal from the QD at 880 nm CW excitation. For the experiments, we use the negative trion transition at 730 mV.}
\label{fig:QDCharacterization}
\end{figure}
\begin{figure}[H]
    \centering
    \includegraphics[width=1\linewidth]{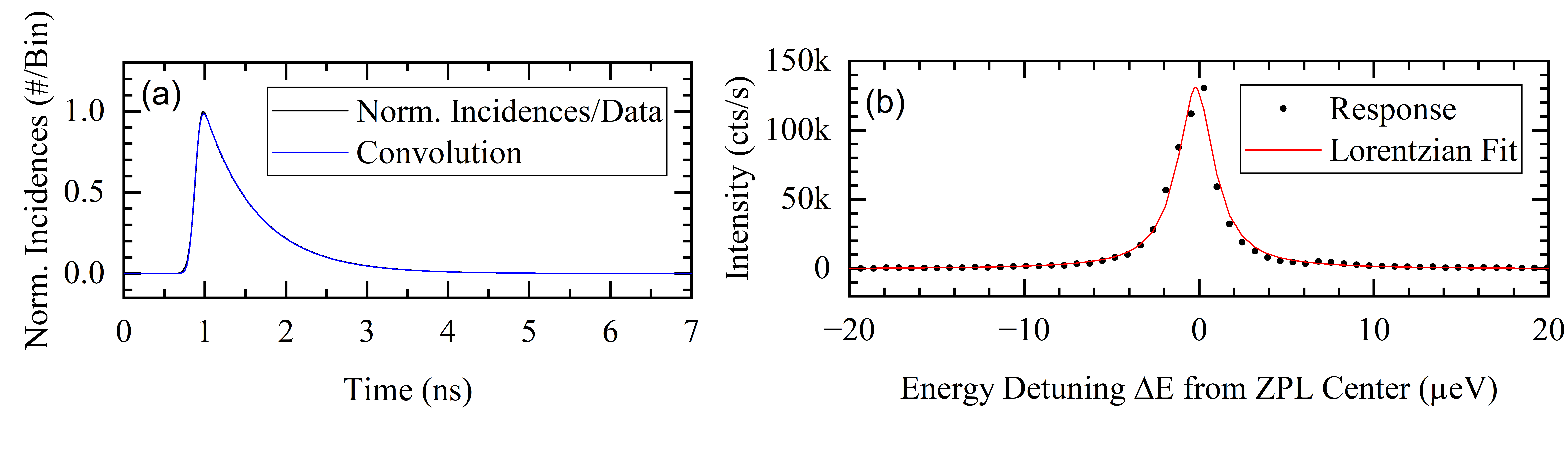}
    \hfill
\caption{(a) Determination of the radiative lifetime according to Eq. (\ref{eq:RadiativeLifetime}). (b) Response to the resonant excitation with a CW laser at 50 pW to determine the according linewidth from the Lorentzian emission, i.e. here the zero-phonon line (ZPL). The Fourier-transform limited linewidth would be 1 $\mu$eV.}
\label{fig:FTL}
\end{figure}

\section{Theoretical framework}

Here we provide the theoretical background needed to understand the blinking models used in this work.

\subsection{Blinking mechanisms}\label{Blinking mechanisms}
One ubiquitous source of fluctuations in the photon flux emitted by semiconductor
quantum emitters is \textit{blinking}. This process can be represented as a
stochastic modulation of the intrinsic intensity $I_0(t)$ by a classical random variable $b(t)$. In this section we will derive the relevant expressions of the blinking contributions to photon statistics for the three blinking mechanisms considered in this work to generate artificial intensity fluctuations.\\\\  
For blinking, the classical random variable $b(t)$ takes the values 1 and 0, corresponding to
ON and OFF states, respectively. If the intrinsic radiative dynamics and the
blinking process occur on well-separated timescales, the measured intensity can
be written as
\begin{equation}
    I_{\mathrm{meas}}(t)=I_0(t)b(t).
\end{equation}

The normalized second-order correlation of the measured emission is then
\begin{equation}
g_{\mathrm{meas}}^{(2)}(\tau)
=
\frac{
\left\langle I_{\mathrm{meas}}(t)
I_{\mathrm{meas}}(t+\tau)\right\rangle
}{
\left\langle I_{\mathrm{meas}}(t)\right\rangle
\left\langle I_{\mathrm{meas}}(t+\tau)\right\rangle
}.
\end{equation}
Assuming that the blinking dynamics are statistically independent of the
intrinsic photon statistics of the emitter, the numerator factorizes as 
$\left\langle
I_{\mathrm{meas}}(t)I_{\mathrm{meas}}(t+\tau)
\right\rangle
=
\left\langle I_0(t)I_0(t+\tau)\right\rangle
\left\langle b(t)b(t+\tau)\right\rangle$,
while $\left\langle I_{\mathrm{meas}}(t)\right\rangle
=
\left\langle I_0(t)\right\rangle
\left\langle b(t)\right\rangle$. 
Consequently, $g_{\mathrm{meas}}^{(2)}(\tau)$ can be written as,  
\begin{equation}
g_{\mathrm{meas}}^{(2)}(\tau)
=
g_0^{(2)}(\tau)\mathcal{B}(\tau),
\end{equation}
where
\begin{equation}
\mathcal{B}(\tau)
=
\frac{
\left\langle b(t)b(t+\tau)\right\rangle
}{
\left\langle b(t)\right\rangle
\left\langle b(t+\tau)\right\rangle
}
\end{equation}
is the normalized autocorrelation of the blinking process. Thus,
\(g_0^{(2)}(\tau)\), as defined on Eq. \ref{eq:g2tau_glauber} of main text, contains the intrinsic photon statistics, whereas
\(\mathcal{B}(\tau)\) describes the additional correlations caused by the slowly-varying
intensity modulation. For a stationary binary process,
\(\langle b(t)\rangle=\langle b(t+\tau)\rangle=P_{\mathrm{on}}\). Moreover,
because \(b(t)b(t+\tau)=1\) only when the emitter is ON at both times,
\begin{align}
\left\langle b(t)b(t+\tau)\right\rangle
&=
\Pr[b(t)=1,b(t+\tau)=1]
\\
&=
P_{\mathrm{on}}
\Pr[b(t+\tau)=1\mid b(t)=1].
\end{align}
Defining the conditional probability
$P_{11}(\tau)
\equiv
\Pr[b(t+\tau)=1\mid b(t)=1],$
the blinking correlation becomes
\begin{equation}\label{BFromCond}
\mathcal{B}(\tau)=\frac{P_{11}(\tau)}{P_{\mathrm{on}}}.
\end{equation}
This expression has a direct physical interpretation; it compares the
probability that the emitter remains in, or returns to, the ON state after a
delay $\tau$, given that it was initially ON,$P_{11}$, with the unconditional
probability $P_\mathrm{on}$ of finding it ON in the first place.
At long delays, the two states become statistically
independent, so \(P_{11}(\infty)=P_{\mathrm{on}}\) and
\begin{equation}
\mathcal{B}(\infty)=1.
\end{equation}
At zero delay, \(P_{11}(0)=1\), giving
\begin{equation}\label{B_zero}
\mathcal{B}(0)=\frac{1}{P_{\mathrm{on}}}.
\end{equation}
The blinking contrast can therefore be expressed through the ratio
\begin{equation}
r = \frac{P_{\mathrm{off}}}{P_{\mathrm{on}}}
=\frac{1-P_{\mathrm{on}}}{P_{\mathrm{on}}}
=\mathcal{B}(0)-1.
\end{equation}

In a pulsed correlation histogram, the peak centered at
\(\tau_n=nT_{\mathrm{rep}}\) counts photon emitted following pulses
separated by \(n\) repetition periods. Its integrated area is therefore
proportional to the joint probability that the emitter is ON at the two pulse
times,
\begin{equation}
S_{\mathrm{env}}(\tau_n)
=
K\left\langle b(t)b(t+\tau_n)\right\rangle,
\end{equation}
where $K$ is a proportionality constant that contains delay-independent experimental factors. At delays
much longer than the blinking correlation time, 
\begin{equation}
C \equiv  S_{\mathrm{env}}(\infty)
=
K P_{\mathrm{on}}^2.
\end{equation}
It follows that
\begin{equation}
\frac{S_{\mathrm{env}}(\tau_n)}{C}
=
\frac{\left\langle b(t)b(t+\tau_n)\right\rangle}
{P_{\mathrm{on}}^2}
=
\mathcal{B}(\tau_n),
\end{equation}
or equivalently,
\begin{equation}\label{S_env_B_connection}
S_{\mathrm{env}}(\tau_n)=C\mathcal{B}(\tau_n).
\end{equation}
Since $\mathcal{B}(\infty) =  1$, the peak envelope can be separated into a constant baseline $C$ and a time-dependent contribution
\begin{equation}\label{Senv decomposed}
S_{\mathrm{env}}(\tau_n)
=C\mathcal{B}(\tau_n)
=C+\Delta S_{\mathrm{env}}(\tau_n),
\end{equation}
where 
$\Delta S_{\mathrm{env}}(\tau)
= C\left[\mathcal{B}(\tau)-1\right]$. 
Here, $\Delta S_{\mathrm{env}}(\tau)$ is understood as the excess area caused by blinking, and vanishes at long delays.\\\\
The three stochastic blinking mechanisms considered in this work are Random Telegraph Noise (RTN), Dark-States and Powerlaw. In the following, we derive the correlation
envelope associated with each mechanism and connect its parameters to the
blinking ratio $r$ and to the envelope-based estimator of $g^{(2)}(0)$.

\subsubsection{Random Telegraph Noise}
Random telegraph noise (RTN), otherwise known as \textit{popcorn noise}, describes a two-state Markov process with constant transition rates $k_{\mathrm{off}}$ from ON to OFF and $k_{\mathrm{on}}$ from OFF to ON \cite{Efros.1997}. The corresponding dwell-time probability densities are
\begin{equation}
\psi_{\mathrm{on}}(t)
=
k_{\mathrm{off}}e^{-k_{\mathrm{off}}t},
\qquad
\psi_{\mathrm{off}}(t)
=
k_{\mathrm{on}}e^{-k_{\mathrm{on}}t},
\end{equation}
with mean dwell times $\tau_{\mathrm{on}}=1/k_{\mathrm{off}}$ and $\tau_{\mathrm{off}}=1/k_{\mathrm{on}}$. The stationary occupation probabilities $P_{\mathrm{on/off}}$ and characteristic correlation time $\tau_b$ are \cite{Gaignard.2025}
\begin{equation}
P_{\mathrm{on}}
=
\frac{\tau_{\mathrm{on}}}{\tau_{\mathrm{on}}+\tau_{\mathrm{off}}},
\qquad
P_{\mathrm{off}}
=
\frac{\tau_{\mathrm{off}}}{\tau_{\mathrm{on}}+\tau_{\mathrm{off}}},
\qquad
\tau_b
=
\frac{1}{k_{\mathrm{on}}+k_{\mathrm{off}}}
=
\frac{\tau_{\mathrm{on}}\tau_{\mathrm{off}}}{\tau_{\mathrm{on}}+\tau_{\mathrm{off}}}.
\end{equation}
The dynamical properties of the RTN process are described by the differential equation for the time dependent probability $p_\mathrm{on}(t)$, 
\begin{align}
    \frac{dp_\mathrm{on}}{dt} = -k_\mathrm{off}p_\mathrm{on}+k_\mathrm{on}(1-p_\mathrm{on}).
\end{align}
With the initial condition $p_\mathrm{on}(0)=1$,
the solution for the differential equation for the ON state is
\begin{align}
    p_\mathrm{on}(t) = P_\mathrm{on} + (1-P_\mathrm{on})e^{-t/\tau_b}.
\end{align}
Given that the emitter is initially ON, the conditional probability follows the same differential equation as $p_\mathrm{on}(t)$. Therefore, the expression defined in Eq.~\eqref{BFromCond} gives
\begin{equation}
P_{11}(\tau)
=
P_{\mathrm{on}}
+
P_{\mathrm{off}}e^{-\tau/\tau_b}.
\end{equation}
Thus, the normalized blinking correlation is
\begin{equation}
\mathcal{B}_{\mathrm{RTN}}(\tau)
=
\frac{P_{11}(\tau)}{P_{\mathrm{on}}}
=
1+r e^{-|\tau|/\tau_b},
\qquad
r=\frac{P_{\mathrm{off}}}{P_{\mathrm{on}}}.
\end{equation}
Using Eq.~\eqref{S_env_B_connection}, the corresponding peak envelope becomes
\begin{equation}\label{eq:S_env_RTN}
S_{\mathrm{env}}^{(\mathrm{RTN})}(\tau)
=
C+A e^{-|\tau|/\tau_b},
\end{equation}
where $A=Cr$. Consequently, the blinking ratio is obtained directly from the fitted envelope as
\begin{equation}
r=\frac{A}{C}.
\end{equation}
Therefore, the functional form of the envelope estimator for the RTN process, using equations (\ref{eq:g2estimators}d) and (\ref{eq:S_env_RTN}) gives 
\begin{align}
    g^{(2)}_\mathrm{env}(0) = \frac{A_0}{S_{\mathrm{env}}^{(\mathrm{RTN})}(0)} =  \frac{A_0}{A + C}. 
\end{align}
We finalize this section by giving an example of the integrated peak area dependence of the RTN process, for different intermittency ratios $r$. The results are shown in Fig. \ref{fig:AppRTNrfactor}, where stars indicate the integrated peak at zero time delay, while dashed lines mark the fit given by Eq. (\ref{eq:S_env_RTN}). From this behavior, it can be seen that for large enough values of $r$, the estimator $g^{(2)}_\mathrm{base} = A_0/S_{\mathrm{env}}^{(\mathrm{RTN})}(\infty) = A_0/C$ yields bunching, since $A_0>C$, giving the opposite nature to the intrinsic non-blinking $g^{(2)}(0)$.

\begin{figure}[h!]
    \centering
    \includegraphics[width=0.8\linewidth]{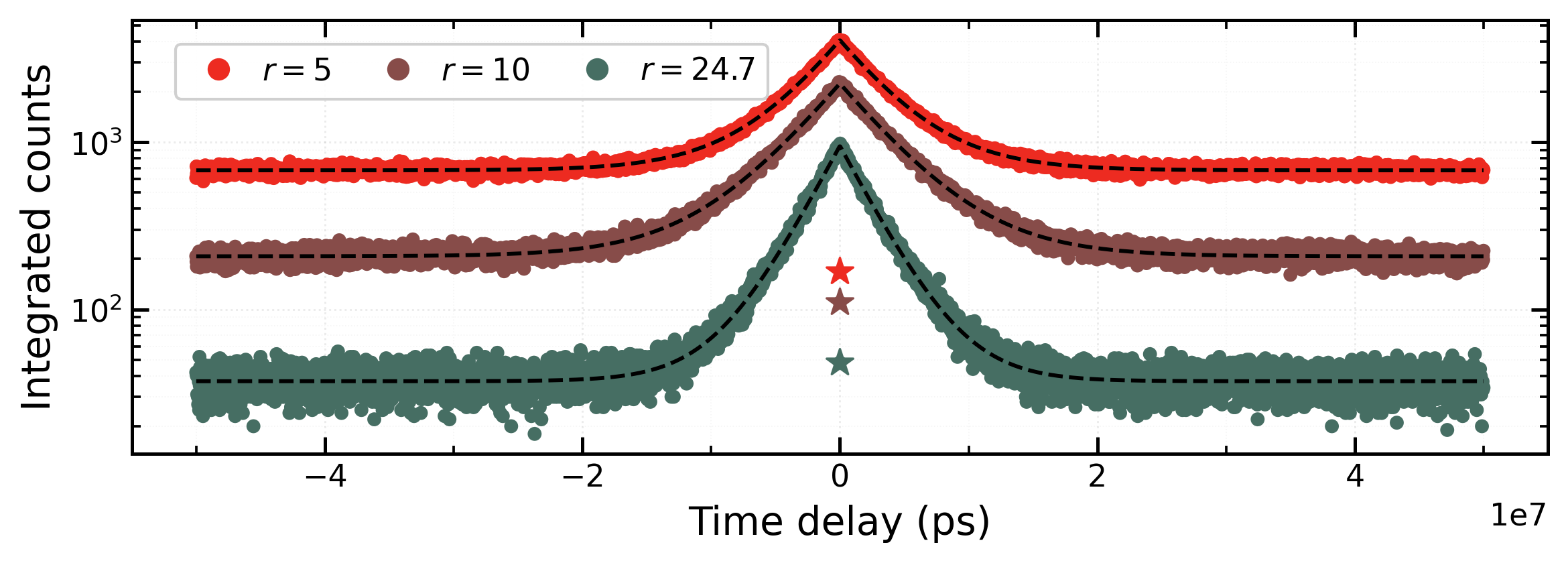}
    \caption{Bidirectional histogram as a function of the time delay obtained by applying an RTN waveform to the raw time-tagged photon streams for different choices of $r$. The stars indicate the integrated peak at zero time delay. The general trend can be found in the main text, Fig. \ref{fig:g2RTNblinking}.}
    \label{fig:AppRTNrfactor}
\end{figure}

\subsubsection{Dark states model}
An extension to the simple RTN model for blinking consists of considering the coupling of a single bright state $B$, where the emission is ON, to N non-radiative dark states $D_i$ \cite{Davanco.2014}. The stochastic variable $b(t)$ is therefore,
\begin{equation}
b(t) = \begin{cases} 
1, & \text{for state } B, \\ 
0, & \text{for any dark state } D_i. 
\end{cases}
\end{equation}
To describe this system, a multilevel rate equation for the bright ($p_B$) and dark ($p_{D_i}$) states probabilities, can be built as
\begin{align}\label{eq:rateModelDS}
    \frac{dp_B}{dt} &= -\sum_{i}^{N}u_ip_B\,+\,\sum_{i}^{N}d_ip_{D_i}\\\nonumber
    \frac{dp_{D_i}}{dt} &= u_ip_B\,-d_ip_{D_i}
\end{align}
where the transition rates from the state $B(D_i)$ to $D_i(B)$ is given by $u_i(d_i)$, The stationary condition for the rate equation yields $P_Bu_i = P_{D_i}d_i$, where $P_{B(D_i)} = \lim_{t \to \infty}p_{B(D_i)}(t)$ are the stationary probabilities. Since such probabilities must fulfill the relation $P_B + \sum_{i}P_{D_i} = 1$, then one obtains
\begin{align}\label{eq:PBStationaryDS}
    P_B = \frac{1}{1+\sum_i u_i/d_i}.
\end{align}
Given that the only radiative state is $B$, then $P_B = P_\mathrm{on}$, while $P_\mathrm{off} = \sum_i P_{D_i} = 1-P_B$. Therefore, the blinking ratio for the dark states can be found as
\begin{align}
    r = \frac{P_\mathrm{off}}{P_\mathrm{on}} = \sum_{i}^N \frac{u_i}{d_i}.
\end{align}
Now, to find the functional form of $S_\mathrm{env}(\tau)$ for the Dark states model, we apply the Laplace transform ($\mathcal{L}\lbrace f(t)\rbrace = \tilde{f}(s)$) to Eq. \ref{eq:rateModelDS}, with the initial condition for the system to be found in the Bright state, i.e., $p_B(0)=1, \, p_{D_i}(0) = 0$. Thus, the transformed rate equations become
\begin{align}
    \mathcal{L}\left\lbrace \frac{dp_B}{dt}\right\rbrace & = s\tilde{p}_B(s)-1 = -\tilde{p}_B(s)\sum_{i}^Nu_i+\sum_{i}^Nd_i\tilde{p}_{D_i}(s)\\\nonumber
\mathcal{L}\left\lbrace \frac{dp_{D_i}}{dt}\right\rbrace &= s\tilde{p}_{D_i}(s) = \tilde{p}_B(s)u_i-d_i\tilde{p}_{D_i}(s),   
\end{align}
where the Laplace transform property for derivatives $\mathcal{L}\lbrace df/dt\rbrace = s\tilde{f}(s)-f(0)$ was applied. Furthermore, after solving for $\tilde{p}_{D_i}(s)$ in the second equation and substituting this result into the first, gives
\begin{align}
    \tilde{p}_B(s) = \frac{1}{s+\sum_{i}u_i-\sum_i \frac{u_i d_i}{s+d_i}}. 
\end{align}
To further exemplify, consider the simpler case of a two Dark state model, i.e., one bright state coupled to two dark states. Upon expansion, the previous equation results in
\begin{align}
    \tilde{p}_{B}(s) = \frac{(s+d_1)(s+d_2)}{(s+u_1+u_2)(s+d_1)(s+d_2) - u_1d_1(s+d_2)-u_2d_2(s+d_1)}.
\end{align}
Applying the stationary probability (Eq.\ref{eq:PBStationaryDS}) for this model, $P_B = d_1d_2/K_2$, where $K_2 = d_1d_2+u_1d_2+u_2d_1$, and 
defining $S_{2} = u_1+u_2+d_1+d_2$ yields,
\begin{align}
    \tilde{p}_{B}(s) = \frac{(s+d_1)(s+d_2)}{s(s^2+ s S_2 + K_2)}.
\end{align}
Simplifying the quadratic term in the denominator, and employing partial fraction decomposition gives
\begin{align}
    \tilde{p}_B(s) = \frac{P_B}{s}+\frac{\Lambda_+}{s+\lambda_+}+\frac{\Lambda_-}{s+\lambda_-},
\end{align}
where $\lambda_{\pm} = \left(-S_2\pm\sqrt{S_{2}^2-4K_2}\right)/2$, and $\Lambda_\pm$ are the residuals of $\tilde{p}_B(s)$ at the poles $\lambda_\pm$. Finally, employing the inverse Laplace transform gives
\begin{align}
    p_B(t) = P_B + \Lambda_+e^{-\lambda_+ t} + \Lambda_-e^{-\lambda_-t}.
\end{align}
This solution is equivalent to that for the conditional probability $P_{11}(\tau)$ within the already imposed initial conditions $p_B(0)=1, p_{D_i} = 0$. Therefore, the blinking correlation (Eq.\ref{BFromCond}) for the two Dark state model becomes,
\begin{align}
    \mathcal{B}_{\mathrm{2DS}}(\tau) = 1 + h_+ e^{-\lambda_+ |\tau|} + h_- e^{-\lambda_- |\tau|}
\end{align}
with $h_{\pm} = \Lambda_\pm/P_B$. From this result, we can extrapolate to a general $N$ dark state model. Therefore, the corresponding peak envelope for the Dark states model becomes
\begin{align}\label{eq:S_env_DS}
    S_{\mathrm{env}}^{\mathrm{DS}}(\tau) = C + \sum_{i}^{N}A_i e^{-|\tau|/\tau_i},
\end{align}
where $\tau_i = \lambda_{i}^{-1}$, $A_i = Ch_i$. The blinking factor extracted from the fit is
$r = \sum_iA_i/C$. 
Therefore, the functional form of the envelope estimator for the DS process, using equations (\ref{eq:g2estimators}d) and (\ref{eq:S_env_DS}) gives 
\begin{align}
    g^{(2)}_\mathrm{env}(0) = \frac{A_0}{S_{\mathrm{env}}^{(\mathrm{DS})}(0)} =  \frac{A_0}{\sum_iA_i + C}. 
\end{align}
\\\\
As a final note, it can also be seen that choosing $N=1$ in this model reproduces the result of Eq.(\ref{eq:S_env_RTN}), upon the connections $u_1 = k_\mathrm{off}$, $d_1 = k_\mathrm{on}$, $\lambda_1 = u_1+ d_1 = \tau_{b}^{-1}$ and $h_1 = u_1/d_1 = P_\mathrm{off}/P_\mathrm{on}$.

\subsubsection{Powerlaw}
In some cases, blinking cannot be described via a Markovian process. The presence of multiple electron traps in the environment that surrounds the QD can lead to tunneling and diffusive processes, resulting in a time-dependent or fluctuating trapping and de-trapping rates \cite{Frantsuzov.2008}. A suitable model for this kind of processes is known as \emph{Powerlaw}. Within suitable assumptions \cite{Shimizu.2001}, the on/off probabilities will follow the law
\begin{equation}
    \psi_\text{on/off}(t) \propto t^{-\alpha_\text{on/off}}.
\end{equation}
The exponents $\alpha_\text{on/off}$ fully determine the blinkling mechanism and they usually range between 1.2 and 2.0.\\\\
Since the system has memory of its past history, the probability that it was ON at time $t$ and that it is still ON at time $t + \tau$ are \emph{not} independent of each other. Furthermore, we assume that every time a transition occurs, the system forgets about its previous state. Thus, we need to use the so-called \emph{renewal theory} \cite{feller1968introduction} to describe this process fully. We start by introducing the probability that the system has a transition ON$\to$OFF at time $t$ via the normalized probability distribution
\begin{align}
    \psi(t > 0) = \frac{\alpha - 1}{\tau_b}\Big( 1 + \frac{t}{\tau_b} \Big)^{-\alpha} \propto \frac{1}{t^\alpha} \qquad \alpha > 1.
\end{align}
From now on, we drop the label "on" to simplify the equations. The short time-scale cutoff $\tau_b$ ensures that the probability distribution follows the Powerlaw and is analytical around $t = 0$.\\\\
Now, the probability that the system stays ON at time $t + \tau$ given the knowledge that it was at time $t$, needs to be calculated. Here, it's important to distinguish between two cases: the system did not have any transition to the OFF state in the interval $(t,t + \tau)$, or the system had one (or more) transitions within the same time interval. In the former case, the probability can be easily calculated as

\begin{align}
    \Psi(\tau) = 1 - \int_0^\tau\text{d} t\psi(t) \implies \tilde{\Psi}(s) = \frac{1 - \tilde{\psi}(s)}{s},
\end{align}
where, again, the tilde symbolizes the application of the Laplace transform. If the system had multiple transitions before being ON again at time $t + \tau$, the probability will be given by the convolution

\begin{align}
    \int\text{d}t' N(t' + \tau)\Psi(t').
\end{align}
$N(t)$ indicates the probability that the system had one or more transitions from 0 to $t$. Using the Laplace transform in the context of renewal theory, it is possible to rewrite the last integral as
\begin{align}
    \tilde{N}(s)\tilde{\Psi}(s) = \frac{\tilde{\psi}(s)}{1 - \tilde{\psi}(s)}\tilde{\Psi}(s), \qquad
    N(s) = \sum_{n=1}^\infty(\tilde{\psi}(s))^n.
\end{align}
Notice that since each transition is independent of the others, the probability of having $n$ transitions in a given time interval is equal to $(\tilde{\psi}(s))^n$ (that corresponds to the $n$-th order convolution of $\psi$ in the time domain). The total probability that the system is ON from time $t$ to time $t + \tau$, in the complex frequency space, will be given by
\begin{align}
    \tilde{S}(s) = \tilde{\Psi}(s) + \tilde{N}(s)\tilde{\Psi}(s) = \dfrac{1}{s\Big(1 - \tilde{\psi}(s)\Big)}\sim \frac{1}{s\,C(s\tau_b)^{\alpha-1}} \qquad s\tau_b\ll1.
\end{align}
$C$ comes from the Taylor expansion of the denominator. By applying the inverse Laplace transform in the long time-scale approximation ($s\tau_b\ll1 \iff \tau\gg\tau_b$), it is possible to see that 

\begin{align}
    S(\tau) \propto \Big(\frac{\tau}{\tau_b}\Big)^{\alpha - 1}.
\end{align}

Instead, in the limit of the short time-scale ($s\tau_b\gg1 \iff \tau\ll\tau_b$), $S(\tau)$ must be a constant. These two limits suggest that the envelope function in the time domain should be of the form
\begin{align}\label{eq:S_env_PL}
    S_\mathrm{env}^{\mathrm{PL}}(\tau) = C + A\left(1+\frac{|\tau|}{\tau_b}\right)^{-\beta} \qquad \beta = 1 - \alpha.
\end{align}
The envelope is a function of the modulus of the delay time $\tau$ thanks to the symmetry properties of the second order auto-correlation function. Therefore, the functional form of the envelope estimator for the PL process, using equations (\ref{eq:g2estimators}d) and (\ref{eq:S_env_PL}) gives 
\begin{align}
    g^{(2)}_\mathrm{env}(0) = \frac{A_0}{S_{\mathrm{env}}^{(\mathrm{PL})}(0)} =  \frac{A_0}{A + C}. 
\end{align}\\\\
\paragraph{Cutoffs} Notice that $\psi(t)$ we introduced is an \emph{improper} probability distribution even with the short time-scale cutoff $\tau_b$. In fact, the first and second moments of the distribution are not guaranteed to converge for any $\alpha > 1$. In order to let $\alpha$ (or equivalently $\beta$) be a free fit parameter, the introduction of a cutoff at long time-scales is necessary. If $1 < \alpha < 2$ ($-1 < \beta < 0$) both $\langle t \rangle$ and $\langle t^2 \rangle$ diverge without the application of a cutoff. The convergence first of $\langle t \rangle$, and then of $\langle t^2 \rangle$ is ensured by increasing $\alpha$, but it depends on the system.

\subsection{Effective correlation function and blinking-induced vacuum admixture}\label{app:grunwaldBlink}

One of the main consequences of blinking is the reduction of the actual ON state fraction of an emitter. Conversely, this can be understood as an additional vacuum contribution to the measured state, which alters the observed photon statistics. In particular, Grünwald \cite{Grunwald.2019} analyzes the increase of the normalized second-order correlation caused by a vacuum admixture and defines an effective metric that removes this contribution. Here, we connect Grünwald's vacuum-corrected second-order correlation function, $g^{(2)}_{\mathrm{G}}(0)$, with our envelope estimator, $g^{(2)}_{\mathrm{env}}(0)$.\\\\
To begin, Grünwald considers a quantum state with an incoherent vacuum contribution,
\begin{equation}
\rho_x=x\ket{0}\bra{0}+(1-x)\rho_0,
\end{equation}
where $x<1$ is the vacuum fraction and $\rho_0$ is the corresponding vacuumless state. The measured second-order correlation of the state containing vacuum is
\begin{equation}
g^{(2)}_x(0)
=
\frac{g^{(2)}_0(0)}{1-x}.
\end{equation}
The previous expression indicates that for the experimental conditions where $0\leq x <1$, $g^{(2)}_x(0)$ overestimates the intrinsic photon statisics of the source. Therefore, the vacuum-corrected effective correlation function is then
\begin{equation}
g^{(2)}_{\mathrm{G}}(0)
= (1-x)g^{(2)}_x(0) = 
g^{(2)}_0(0).
\label{eq:Grunwaldcorrectg2}
\end{equation}
This treatment can be directly connected to blinking, since it consists of a two-state process in which each excitation pulse probes either the ON state of the emitter or an OFF state that contributes no photons. The state can therefore be written as
\begin{equation}
\rho_{\mathrm{blink}}
= P_{\mathrm{off}}\ket{0}\bra{0}
+
P_{\mathrm{on}}\rho_{\mathrm{ON}},
\end{equation}
where $P_{\mathrm{on}}$ and $P_{\mathrm{off}}$ are the occupation probabilities of the ON and OFF states. If $\rho_{\mathrm{ON}}$ has no vacuum component of its own, blinking maps exactly onto Grünwald's vacuum admixture through
\begin{equation}
x=P_{\mathrm{off}},
\qquad
1-x=P_{\mathrm{on}}.
\end{equation}
As shown in Eq.~\ref{S_env_B_connection}, for pulsed histograms there is a direct connection between the blinking autocorrelation $B(\tau_n)$ and the integrated side peak area $S_\mathrm{env}(\tau)$. In particular, extrapolating the envelope to zero delay and using Eq.~\ref{B_zero} gives
\begin{equation}
P_{\mathrm{on}}
=
\frac{C}{S_{\mathrm{env}}(0)}.
\end{equation}
Recalling the definition of the baseline estimator from Eq.~\ref{eq:g2estimators}e,
\begin{equation}
g^{(2)}_{\mathrm{base}}(0)
=
\frac{A_0}{C} = \frac{A_0}{S(\infty)},
\end{equation}
and applying Grünwald's vacuum correction from Eq.~\ref{eq:Grunwaldcorrectg2}, we obtain
\begin{equation}
g^{(2)}_{\mathrm{G}}(0)
= P_{\mathrm{on}}g^{(2)}_{\mathrm{base}}(0)=
\frac{C}{S_{\mathrm{env}}(0)}
\frac{A_0}{C}.
\end{equation}
Therefore,
\begin{equation}
g^{(2)}_{\mathrm{G}}(0)
=
\frac{A_0}{S_{\mathrm{env}}(0)}
 = g^{(2)}_{\mathrm{env}}(0).
\end{equation}
Thus, when the ON-state field is vacuumless, the envelope-normalized estimator $g^{(2)}_{\mathrm{env}}$ used in the main text (Eq. \ref{eq:g2estimators}d) corresponds exactly to Grünwald's effective vacuum-corrected correlation function for a blinking source whose OFF state behaves as vacuum. More generally, $\rho_{\mathrm{ON}}$ may itself contain a vacuum component
$x_{\mathrm{ON}}=\bra{0}\rho_{\mathrm{ON}}\ket{0}$. The total vacuum fraction is then
\begin{equation}
x_{\mathrm{tot}}
=
P_{\mathrm{off}}+P_{\mathrm{on}}x_{\mathrm{ON}}.
\end{equation}
In this case, the envelope estimator removes only the additional vacuum contribution caused by blinking and recovers $g^{(2)}_{\mathrm{ON}}(0)$. Grünwald's correction of the total vacuum component instead gives
\begin{equation}
g^{(2)}_{\mathrm{G}}(0)
=
(1-x_{\mathrm{ON}})g^{(2)}_{\mathrm{env}}(0).
\end{equation}

\section{Reduced $\chi^2$}

One of the most common ways to quantify the agreement between a theoretical model and an experimental data set is via the Bayesian maximum likelihood estimation. If the number of observations ($N$) of all possible outcomes of a measurement is large enough, then the posterior distribution of the theoretical parameters given the experimental data will follow the $\chi^2$ distribution with $N - M$ degrees of freedom. Here $M$ defines the number of fitted parameters.

We want to use the $\chi^2$-test to measure the agreement between the central-to-nearest neighbor evaluation of the single-photon purity and the predictions from the models shown above. Since each model has its own number of fit parameters, we would like the result to be less sensitive to these differences. To solve this inconvenience, it is useful to use the reduced $\chi^2$.

When data is properly fitted without over- or underfitting, the reduced $\chi^2$ statistic provides a valid metric for directly comparing how well different models agree with the collected data.

\end{widetext}
\end{document}